\documentclass[aps]{revtex4}
\usepackage{graphicx,color,amsmath}
\begin{document}
\title{Light Strange-beauty Squarks
\footnote{Parallel talk given in the SUSY session at
the International Conference on 
Linear Colliders, LCWS04, 19-23 April 2004, Paris, France,
and in the Flavor and CP session at SUSY 2004, June 17-23, 2004, Tsukuba,
Japan.}
}
\author{Kingman Cheung}
\address{
Department of Physics and NCTS, National Tsing Hua
University, Hsinchu, Taiwan, R.O.C.}
\date{\today}
%\maketitle
\begin{abstract}
%\abstract{
We summarize a study on the production, decay, and detection of
the strange-beauty squark as light as 200 GeV at hadronic and $e^+ e^-$
colliders.  It was motivated by nearly maximal mixing between strange
and beauty squarks.
%}
\end{abstract}
\maketitle

\section{Introduction}
The SM predicts that the mixing-dependent CP violation
measured in $B\to \phi K_S$ should be the same as in
$B\to J/\psi K_S$.
Belle, however, has found an opposite sign in
$B\to \phi K_S$ for two consecutive years~\cite{Belle}.
The current discrepancy with SM
prediction stands at a 3.5$\sigma$ level. 
Although Belle result is not consistent with BaBar, 
the combined result still has a 2.7$\sigma$ deviation from
SM expectation. 
It was pointed out~\cite{CH} that a light
strange-beauty squark, resulted from near-maximal mixing between
the second and third right-handed (RH) squarks,
provides all necessary ingredients
to account for the discrepancy. 
It provides
(1) a large $s$--$b$ flavor mixing, (2) a unique new CP
violating phase, and (3) right-handed dynamics.  The latter is
needed for explaining why similar ``wrong-sign" effects are not
observed in the modes such as $B\to K_S\pi^0$ and $\eta^\prime
K_S$. 
This near-maximal mixing is motivated by 
a combination of Abelian flavor symmetry (AFS) and SUSY.
Such a near-maximal mixing allows
one state to be considerably lighter than the common squark mass
scale $\widetilde m$.  
A detailed study of various $B$ decays
suggested~\cite{CHNphik} that $m_{\widetilde{sb}_1} \sim 200$ GeV
and $m_{\tilde g} \sim 500$ GeV are needed, while the squark mass
scale $\widetilde m$ can be well above 1 TeV.

It is clear that a squark as light as 200 GeV is of great interest
at the Tevatron. One should
independently pursue the search for such a light
$\widetilde{sb}_1$, even if the $B\to \phi K_S$ CP
violation discrepancy evaporates in the next few years. We note
that a strange-beauty squark, carrying $\sim$ 50\% in strange and
beauty flavor, would lead to a weakening of bounds on beauty
squark search based on $b$-tagging.  
The SUSY scenario that we have is that (i) the 
generic soft SUSY scale is at TeV,
(ii) a RH strange-beauty squark as light as 200 GeV,
(iii) a relatively light gluino $\sim 500$ GeV that goes together with
the squark in the gluino-squark loop.
We summarize our work \cite{hou}
on strange-beauty squark-pair production, and 
various decay scenarios of the squark at the Tevatron.

\section{Hadronic Production}

We focus only on
the $2\times 2$ right-handed strange and beauty squarks,
whose mass matrix is given by,
\begin{equation}
{\cal L} = - ( \tilde{s}_R^* \; \tilde{b}_R^* )\;
\left( \begin{array}{ll}
   \widetilde{m}_{22}^2 &  \widetilde{m}_{23}^2 e^{-i\sigma} \\
   \widetilde{m}_{23}^2 e^{i\sigma} & \widetilde{m}_{33}^2 \end{array}
  \right ) \;
\left( \begin{array}{c}
      \tilde{s}_R \\
      \tilde{b}_R \end{array} \right ) \;,
\end{equation}
which can be diagonalized by the transformation
\begin{equation}
\left( \begin{array}{c}
     \tilde{s}_R \\
     \tilde{b}_R \end{array} \right ) = R \,
\left( \begin{array}{c}
     \widetilde{sb}_1 \\
     \widetilde{sb}_2 \end{array} \right ) =
\left( \begin{array}{ll}
    \cos\theta_m& \sin\theta_m\\
    -\sin\theta_me^{i\sigma} & \cos\theta_me^{i\sigma} \end{array} \right ) \;
\left( \begin{array}{c}
     \widetilde{sb}_1 \\
     \widetilde{sb}_2 \end{array} \right ) \;.
\end{equation}
The relevant gluino-quark-squark interactions
in the mass eigenbasis are
\begin{eqnarray}
{\cal L} &=&  -\sqrt{2} g_s T^a_{kj} \left [
 - \overline{\widetilde{g}_a} P_R s_j \widetilde{sb}_{1k}^* \cos\theta_m
 + \overline{\widetilde{g}_a} P_R b_j \widetilde{sb}_{1k}^* \sin\theta_m
   e^{-i\sigma} \nonumber \right .\\
& & \left.
 - \overline{\widetilde{g}_a} P_R s_j \widetilde{sb}_{2k}^* \sin\theta_m
 - \overline{\widetilde{g}_a} P_R b_j \widetilde{sb}_{2k}^* \cos\theta_m
   e^{-i\sigma}
+ {\rm h.c.}
 \right ] \; \nonumber \\
 && -i g_s A^a_\mu T^a_{ij} \,\left( \widetilde{sb}_{1i}^*
 {\stackrel{\leftrightarrow}{\partial}}_\mu \widetilde{sb}_{1j}
 + \widetilde{sb}_{2i}^*
 {\stackrel{\leftrightarrow}{\partial}}_\mu \widetilde{sb}_{2j} \right )
 \nonumber \\
&& + g_s^2 (T^a T^b)_{ij} A^{a\mu} A^b_{\mu}
\left( \widetilde{sb}_{1i}^*  \widetilde{sb}_{1j}
 + \widetilde{sb}_{2i}^* \widetilde{sb}_{2j} \right ) \;.
\end{eqnarray}

The production of the strange-beauty squark can proceed via
\begin{eqnarray}
 q\bar q,\; gg &\to& \widetilde{sb}_1\,  \widetilde{sb}_1^*\;, \nonumber \\
 ss,\ sb,\ bb  &\to& \widetilde{sb}_1\,  \widetilde{sb}_1 \;, \nonumber \\
q\bar q,gg &\to& \tilde{g}\tilde{g}\;\;\; \mbox{followed by gluino decay}, 
\;\;{\rm and} 
  \nonumber \\
sg,\, bg\, &\to& \widetilde{sb}_1 \tilde{g}\;\;\; 
\mbox{followed by gluino decay} \;. \nonumber 
\end{eqnarray}
The gluino so produced will decay
into a strange or beauty quark plus the strange-beauty squark
$\widetilde{sb}_1$.  

\begin{figure}[t!]
\centering
\includegraphics[width=3in]{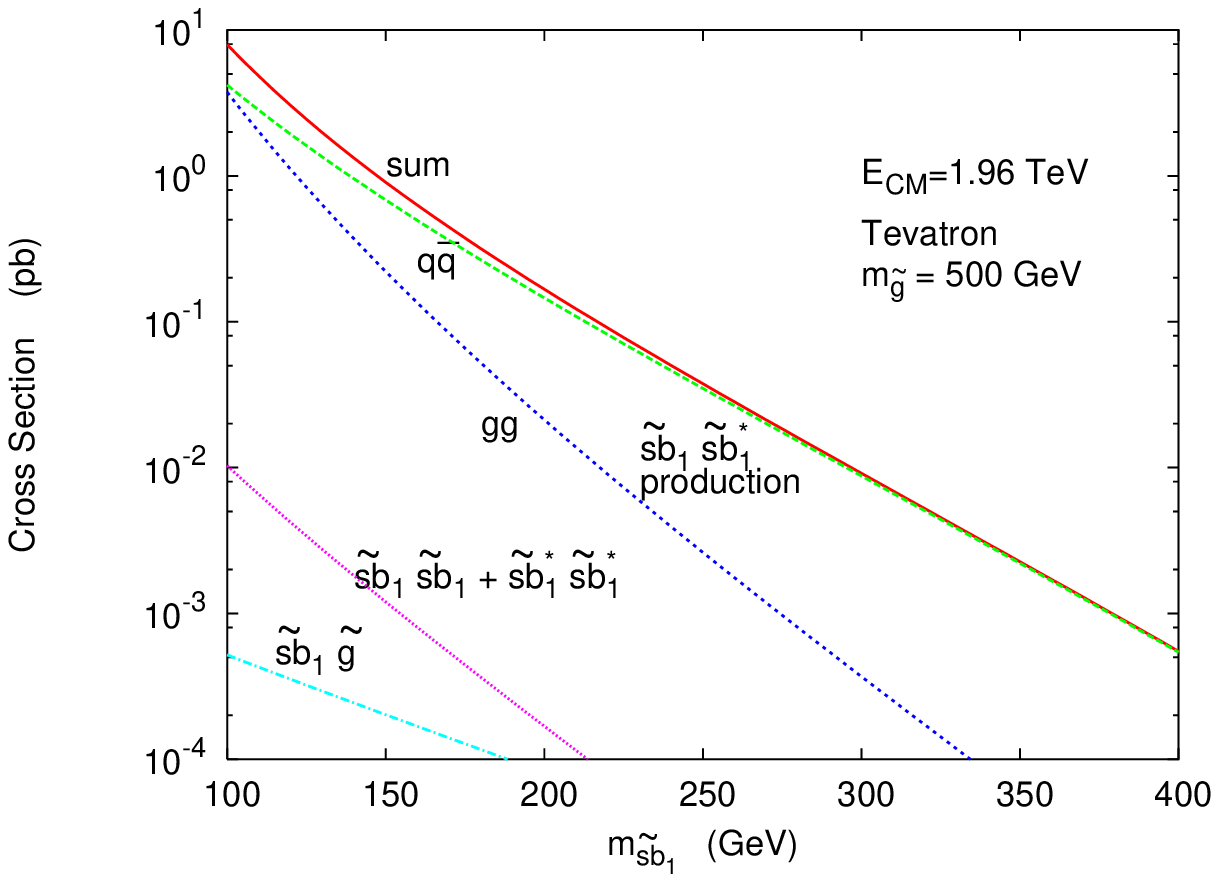}
\includegraphics[width=3in]{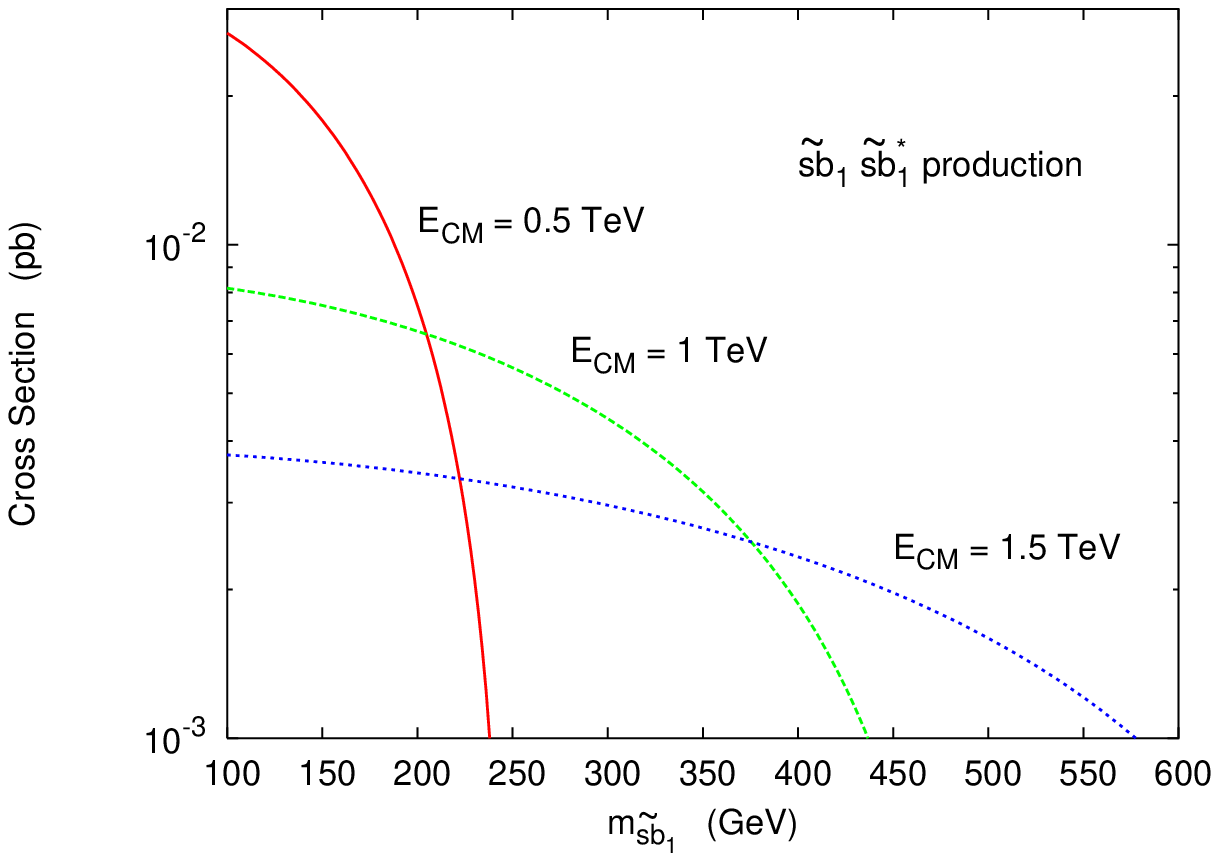}
\caption{\small \label{xs-teva} (a) Total cross section for 
direct production of the $\widetilde{sb}_1 \widetilde{sb}_1^*$ pair
at the Tevatron. (b) Production cross sections for pair production at
$e^+ e^-$ colliders.
}
\end{figure}

Production cross sections for direct $\widetilde{sb}_1
\widetilde{sb}_1^*$ pair production at the Tevatron are shown in
Fig.~\ref{xs-teva}(a).
As expected, the production is dominated by $q\bar q$ annihilation.
For a gluino mass of 500 GeV the gluino-pair cross section is
only a few fb, but, however, the feed-down from gluino 
becomes comparable to direct $\widetilde{sb}_1$ pair production when
$m_{\widetilde{sb}_1} \agt 300$ GeV.  Therefore, 
at high $m_{\widetilde{sb}_1}$ the gluino contribution can help
extending  the sensitivity further.

The situation is different at the LHC, please refer to Ref. \cite{hou}
for more details.  Gluino pair production and squark pair production
cross sections are both above 10 pb for
$m_{\widetilde{sb}_1} = 200$ GeV and $m_{\tilde g} = 500$ GeV, and
both contributions have to be taken into account.  We also show the 
production cross sections at $e^+ e^-$ colliders in Fig.~\ref{xs-teva}(b).

\section{Decay and detection of the strange-beauty squark}

We  consider three scenarios for the decay of $\widetilde{sb}_1$.

\subsection{Strange-beauty Squark  as the LSP}
When the $\widetilde{sb}_1$ is the LSP and $R$-parity is conserved, 
the $\widetilde{sb}_1 \widetilde{sb}_1^*$ pair so
produced will hadronize into color-neutral hadrons by combining
with some light quarks. 
Such objects are strongly-interacting massive particles,
electrically either neutral or charged.  If the hadron is
electrically neutral, it will pass through the tracker with little
trace. 
However, if it is electrically charged 
it  will undergo ionization
energy loss in the central tracking system, hence behaves like a
``heavy muon''. 
We did a similar analysis as CDF search for stable charged particles 
\cite{cdf-stable} with the following selective cuts:
$
p_T(\widetilde{sb}_1) > 20 \;{\rm GeV}\;, |y(\widetilde{sb}_1)|<2.0\;,
 0.25 < \beta\gamma < 0.85\;.
$
In Ref.~\cite{hou} we have shown the cross sections from direct
$\widetilde{sb}_1$ pair production with all the above
acceptance cuts, for detecting 1 massive
stable charged particle (MCP), 2 MCPs, or at least 1 MCPs in the
final state.  
The sensitivity can reach up to almost 
$m_{\widetilde{sb}_1}\simeq300$ GeV with an integrated luminosity 
of 2 fb$^{-1}$.

\subsection{$\widetilde{sb}_1$ as LSP but $R$-parity is violated}
In this case the $\widetilde{sb}_1$ pair so produced will decay
via the $R$-parity violating couplings $\lambda' LQD^c$ or
$\lambda^{''} U^c D^c D^c$.  
For simplicity we only consider $\lambda'_{ii3},\ \lambda'_{ii2}$ with
$i=1,\ 2$. The strange-beauty squark will decay into $e^-u$ or
$\mu^- c$, and thus behaves like
scalar leptoquarks of the first or second generation, respectively. 
The most stringent limit on leptoquarks comes from 
a combined analysis that pushes the first
generation leptoquark limit to around 260 GeV \cite{lepto-limit}.
Therefore, we believe that the limit
that can be reached at the end of Run II (2fb$^{-1}$) is very 
likely above 300 GeV.  With an order more luminosity, the limit 
may be able to reach 350 GeV.

\subsection{$\widetilde{sb}_1$ is the NLSP}
In this case the $\widetilde{sb}_1$ so produced will decay into a
strange or beauty quark plus the neutralino or the gravitino, depending
on the SUSY breaking models.
There are 2 quark jets in the final state of
$\widetilde{sb}_1\widetilde{sb}_1^*$ pair production, each of them
either strange or beauty flavored, and with large missing energy
due to the neutralinos or gravitinos.  We performed an analysis 
with the following selective cuts and $b$-tagging and mistag efficiencies:
$
p_{Tj} > 15 \;{\rm GeV}\;,  |\eta_j|< 2.0\;, 
\not\!{p}_T>40 \;{\rm GeV}, \epsilon_{btag} = 0.6\;,  \epsilon_{mis} 
= 0.05 \;.
$
Note that the branching ratio of the strange-beauty squark into a
$b$ quark scales as $\sin^2\theta_m$.
With an integrated luminosity of 2 fb$^{-1}$ the
sensitivity is around 300 GeV, if $\sin^2\theta_m\agt 0.5$. If the
integrated luminosity can go up to 20 fb$^{-1}$, then the
sensitivity increases to 350 GeV.
We also emphasize that the double-tag vs single-tag ratio contains
information on $\sin^2\theta_m$, while their sum, when compared with
the standard $\tilde b$ squark pair production, provides
additional consistency check on  cross section vs mass. 

In conclusion, the recent possible CP violation discrepancy in
$B\to \phi K_S$ decay suggests the possibility of a light
strange-beauty squark $\widetilde{sb}_1$ that carries both strange
and beauty flavors. Such an unusual squark can be searched for at
the Tevatron Run II, with the precaution that $\widetilde{sb}_1$
can decay into a beauty or strange quark, and the standard
$\tilde b$ search should be broadened. Discovery up to 300 GeV is
not a problem, and anomalous behavior in both production cross
sections and the single versus double tag ratio may provide
confirming evidence for the strange-beauty squark.

\section*{Acknowledgment}
I am grateful to Wei-Shu Hou for collaboration. 
This research was supported in part by
the National Science Council of Taiwan R.O.C. under grant no.
NSC 92-2112-M-007-053-.


\begin{thebibliography}{99}
%
\bibitem{Belle}
Belle Coll., Phys.\ Rev.\ D {\bf 67}, 031102 (2003);
Phys. Rev. Lett.  {\bf 91}, 261602 (2003).
%%
\bibitem{CH}
C.K. Chua and W.S. Hou, Phys. Rev. Lett. {\bf 86}, 2728 (2001).
%%CITATION = HEP-PH 0005015;%%;
\bibitem{CHNphik}
C.K. Chua, W.S. Hou and M. Nagashima, Phys. Rev. Lett. {\bf 92}, 201803 (2004).
%
%
\bibitem{hou}
K. Cheung and W.S. Hou, hep-ph/0404041, to appear in PRD.

\bibitem{cdf-stable}
CDF Coll., Phys. Rev. Lett. {\bf 90}, 131801 (2003).

\bibitem{lepto-limit}
The latest preliminary results from D\O\ is available at {\tt
http://www-d0.fnal.gov/Run2Physics/np/}.

\end{thebibliography}
\end{document}